\documentclass{nature}
\newcommand {\gtrsim} {\ {\raise-.5ex\hbox{$\buildrel>\over\sim$}}\ }
\newcommand {\ltrsim} {\ {\raise-.5ex\hbox{$\buildrel<\over\sim$}}\ }
\usepackage{psfig}
\usepackage{graphicx}

\title{Two stellar-mass black holes in the globular cluster M22}

\author{Jay Strader$^{1,2}$, Laura Chomiuk$^{3,1,2}$, Thomas J.~Maccarone$^4$,  James C.~A.~Miller-Jones$^5$, Anil C.~Seth$^6$}

\begin{document}
\spacing{1}

\maketitle

\begin{affiliations}
\item Department of Physics and Astronomy, Michigan State University, East Lansing, Michigan 48824, USA
\item Harvard-Smithsonian Center for Astrophysics, Cambridge, MA 02138, USA
\item National Radio Astronomy Observatory, P.O. Box O, Socorro, NM 87801, USA
\item School of Physics and Astronomy, University of Southampton, Highfield SO17 IBJ, UK
\item International Centre for Radio Astronomy Research, Curtin University, GPO Box U1987, Perth, WA 6845, Australia
\item Department of Physics and Astronomy, University of Utah, Salt Lake City, UT 84112, USA
\end{affiliations}

\begin{abstract}

Hundreds of stellar-mass black holes likely form in a typical globular star cluster, with all but one predicted to be ejected through dynamical interactions\cite{Kul93,Sig93,Por00}. Some observational support for this idea is provided by the lack of X-ray-emitting binary stars comprising one black hole and one other star (``black-hole/X-ray binaries") in Milky Way globular clusters, even though many neutron-star/X-ray binaries are known\cite{Kal04}. Although a few black holes have been seen in globular clusters around other galaxies\cite{Mac07a,Irw10}, the masses of these cannot be determined, and some may be intermediate-mass black holes that form through exotic mechanisms\cite{Por04}. Here we report the presence of two flat-spectrum radio sources in the Milky Way globular cluster M22, and we argue that these objects are black holes of stellar mass (each $\sim 10$--$20$ times more massive than the Sun) 
that are accreting matter. We find a high ratio of radio-to-X-ray flux for these black holes, consistent with the larger predicted masses of black holes in globular clusters compared to those outside\cite{Bel10}. The identification of two black holes in one cluster shows that the ejection of black holes is not as efficient as predicted by most models\cite{Kul93,Sig93,Kal04}, and we argue that M22 may contain a total population of $\sim 5$--$100$ black holes. The large core radius of M22 could arise from heating produced by the black holes\cite{Mac08}.

\end{abstract}

We have obtained very deep radio continuum images of the Milky Way globular cluster M22 (NGC~6656) with the Karl G.~Jansky Very Large Array (VLA). The principal goal of the observations was to search for a possible central intermediate-mass black hole via synchrotron emission from the accretion of intracluster gas; no central source was found\cite{Str12}. However, we serendipitously detected two previously unknown radio continuum sources in the core of the cluster (Fig.~1). We term the sources M22-VLA1 and M22-VLA2. Both sources have flat radio spectra and are unresolved at our $\sim 1''$ resolution. 

The core radius of M22 is uncommonly large for a Milky Way globular cluster: $\sim 1.24$ pc\cite{McL05}. These sources are well inside the cluster core, at projected radii of 0.4 pc and 0.25 pc for M22-VLA1 and  M22-VLA2 respectively. The distance of the next source of comparable flux density is far outside the core, at a projected radius of 2.4 pc.

These sources have no counterparts in shallow archival \emph{Chandra} X-ray imaging. Based on these non-detections, the sources are constrained to have  $L_X \ltrsim 2.2 \times 10^{30}$ erg s$^{-1}$ over 3--9 keV at the distance of M22\cite{Mon04}. The radio luminosities of the sources are $L_R \sim 6\times10^{27}$ erg s$^{-1}$ at 8.4 GHz, assuming flat spectra. Therefore, if the sources are not variable, the limit of radio to X-ray luminosity is: log $L_R/L_X \gtrsim -2.6$.

The radio luminosity, $L_R/L_X$ ratio, and central location of the sources place significant constraints on their nature. The most likely explanation is that both sources are accreting stellar-mass black holes in M22. Other possibilities, all of which we consider unlikely, are discussed in the Supplementary Information. These objects are the first strong candidates for stellar-mass black holes in any Milky Way globular cluster, and the first stellar-mass black holes to be discovered through radio emission rather than via X-rays.\cite{Mac07b}

The radio emission implies that the black holes are actively accreting, and the flat radio spectra are consistent with relatively low accretion rates\cite{Gal05} ($\ltrsim 2$--3\% of the Eddington rate). Because globular clusters have modest amounts of interstellar gas, it is very unlikely that the radio luminosity can be explained by Bondi accretion. Thus the objects cannot be black-hole/black-hole binaries, and instead are probably in binary systems with Roche lobe-overflowing companions. Stellar-mass black holes ($\sim 5$--$100$ $M_{\odot}$) offer the best explanation for the presence of multiple sources close to the cluster center; objects more massive than the average cluster star will sink to the center because of mass segregation.

To look for optical counterparts to the radio sources, we used archival \emph{Hubble Space Telescope} ($HST$) imaging of M22, for which photometric catalogs are available\cite{And08}. Fig.~2 shows that M22-VLA1 is a close match ($0.05''$) to a moderately low-mass ($\sim 0.34$ $M_{\odot}$) main sequence M dwarf in M22 as inferred from standard stellar isochrones (see Supplementary Information for more details). M22-VLA2 is $0.17''$ from the nearest detected star, which is a $\sim 0.62$ $M_{\odot}$ main sequence star. Considering the distribution of stars in the inner $30''$ of the cluster,  the probability of a chance coincidence as close as for M22-VLA1 is only 2\%; for  M22-VLA2 it is 26\%. Thus we consider the optical association for source M22-VLA1 suggestive, but that for M22-VLA2 uncertain. However, for the case of M22-VLA1, there is an additional complication: since the average stellar mass in the core is greater than that of the putative companion, the low-mass main sequence star would likely be exchanged out of the binary in a three-body interaction with another star\cite{Kal04}. On the other hand, because of the low central density of M22\cite{McL05} ($< 10^{4}$ $M_{\odot}$ pc$^{-3}$), a binary with a low-mass companion may survive longer than in a typical globular cluster. Nonetheless, it is possible that both radio sources are associated with low luminosity objects below the detection limit of the $HST$ data, such as white dwarfs.

Stellar-mass black holes with accretion rates below $\sim 2\%$ of the Eddington rate\cite{Mac03} (in the so-called low/hard state) follow an empirical correlation between radio and X-ray luminosity with a scatter of $\sim$ a factor of two\cite{Gal03}. Fig.~3 shows this correlation with the M22 data overplotted. The radio--X-ray relation predicts an X-ray luminosity of $10^{31}$--$10^{32}$ erg sec$^{-1}$ for this radio luminosity\cite{Gal06,Mil11}, above the completeness limit of the archival \emph{Chandra} data.  There are several plausible explanations for this discrepancy. First, there is the possibility of variability. The X-ray data were taken in in 2005, six years earlier than the radio data. Field stellar-mass black holes in the low/hard state show substantial (typically a factor of 2--10) variability in both radio and X-rays\cite{Mil08,Cor06}. Therefore, concurrent radio and X-ray data are necessary for precise constraints on $L_R/L_X$. We found marginal evidence for radio variability in M22-VLA2 on the timescale of a week; more details can be found in the Supplementary Information. Another plausible explanation is that there is larger scatter in the radio--X-ray correlation at very low accretion rates. Only a single known black hole binary has a measured radio luminosity as faint as our sources\cite{Gal06}, and there is evidence that some stellar-mass black holes with low X-ray luminosities may not fall on the correlation\cite{Mil11}. 

An intriguing possibility is that these sources have high values of $L_R/L_X$ because they are more massive than typical stellar-mass black holes in the field. The radio--X-ray correlation for stellar-mass black holes is a special case of a ``fundamental plane" for black hole accretion in the low/hard state that includes the black hole mass as a third parameter\cite{Mer03}. In this relation, more massive black holes have larger values of $L_R/L_X$. If our sources have masses of $\sim 15$--20 $M_{\odot}$ rather than the 5--10 $M_{\odot}$ typical of field stellar-mass black holes\cite{Rem06}, then their X-ray luminosities should be lower than predicted by the correlation in Fig.~3 by a factor of $\sim 2$--3. It is reasonable to expect that black holes in globular clusters will be more massive than those in the field. Field black holes with measured dynamical masses are all in binary systems and were probably affected by mass transfer during a common envelope stage that reduced the mass of the resulting black holes\cite{Tau06}. This need not be the case in globular clusters, since black holes can form as single objects or in wide binaries, and then be exchanged into pre-existing binaries or tidally capture companions due to the high stellar densities\cite{Iva10}. Globular cluster black holes also form at lower metallicity than in the field, leading to less mass loss from the progenitor and thus more massive remnants\cite{Bel10}. 

As mentioned above, the location of stars in a cluster also gives information about their masses. Stellar-mass black holes will mass-segregate to the core of the cluster. This process can be used to roughly estimate their masses by assuming thermalization, for which this relation holds$^{1}$: $m_{BH} / m_{\star} = (r_{c}/r_{BH})^{2}$, where $m_{BH}$ and $r_{BH}$ are the characteristic black hole mass and radius, $m_{\star}$ is the typical stellar mass,  and $r_c$ is the core radius. Assuming $m_{\star} = 1$ $M_{\odot}$ in the segregated cluster core and taking the observed values of $r_c = 1.24$ pc and $r_{BH} = 0.33$ pc, we estimate $m_{BH} \sim 15$ $M_{\odot}$.

The existence of black holes in a low-density globular cluster such as M22 constrains the magnitude of initial velocity kicks received by the black holes at birth. The current central escape velocity of M22\cite{McL05} is $\sim 34$ km s$^{-1}$. This value may have been higher in the past, due to a larger cluster mass and a more compact structure. Nonetheless, the retention of two black holes in a globular cluster with a modest escape velocity implies that the black holes could not have received large natal kicks. Large kicks are inferred for some stellar-mass black holes in the field\cite{Rep12}. Low kick velocities can originate from supernovae if the black hole mass is large, or if the black holes form from direct collapse with no supernovae. In either case, higher black holes masses are favored.

The presence of black holes in a globular cluster can lead to an expansion of the core radius through interactions between black holes and stars. This could explain why M22 has the fifth-largest core radius among luminous ($\gtrsim 2\times 10^{5} L_{\odot}$) Milky Way globular clusters\cite{McL05}. Additional discussion can be found in the Supplementary Information.

Most theoretical models in the literature predict that only a single black hole (or black-hole/black-hole binary) will survive the dynamical processes by which black holes mass-segregate to the cluster center, form an unstable subcluster, and evaporate\cite{Kul93,Sig93,Kal04}. In some cases more than one black hole may temporarily survive for an additional black hole relaxation time ($< 1$ Gyr), if the extra black holes are kicked into orbits outside the core\cite{Sig93}. Additional discussion can be found in the Supplementary Information.

In contrast to these theoretical predictions, M22 contains more than one black hole. In fact, it is possible that more than two black holes are present in M22, either as single black holes or in binary systems that are not undergoing observable mass transfer. Under the uncertain assumption that both of the M22 sources are black hole--white dwarf binaries, published calculations can be used to estimate the fraction of surviving black holes that are actively accreting in present-day globular clusters\cite{Iva10}. Over 10 Gyr, 2--40\% of black holes are expected to become members of binary systems with observable accretion. Our two observed sources thus suggest a total population of $\sim 5$--100 black holes in M22.

\begin{addendum}

\item[Supplementary Information ]is linked to the online version of the paper at www.nature.com/nature.

\item National Radio Astronomy Observatory a facility of the National Science Foundation operated under cooperative agreement by Associated Universities, Inc. L.C. is a Jansky Fellow of National Radio Astronomy Observatory. This work is partially based on observations made with the NASA/ESA Hubble Space Telescope, and obtained from the Hubble Legacy Archive, which is a collaboration between the Space Telescope Science Institute (STScI/NASA), the Space Telescope European Coordinating Facility (ST-ECF/ESA) and the Canadian Astronomy Data Centre (CADC/NRC/CSA).

\item[Author Contributions] J.S. wrote the text. L.C. reduced the data. All authors contributed to the interpretation of the data and commented on the final manuscript.

\item[Competing Interests] The authors declare that they have no competing financial interests.

\item[Correspondence] Correspondence and requests for materials should be addressed to J.~S. (email: strader@pa.msu.edu).

\end{addendum}

\clearpage
\begin{figure}
\centering
\includegraphics[width=5in]{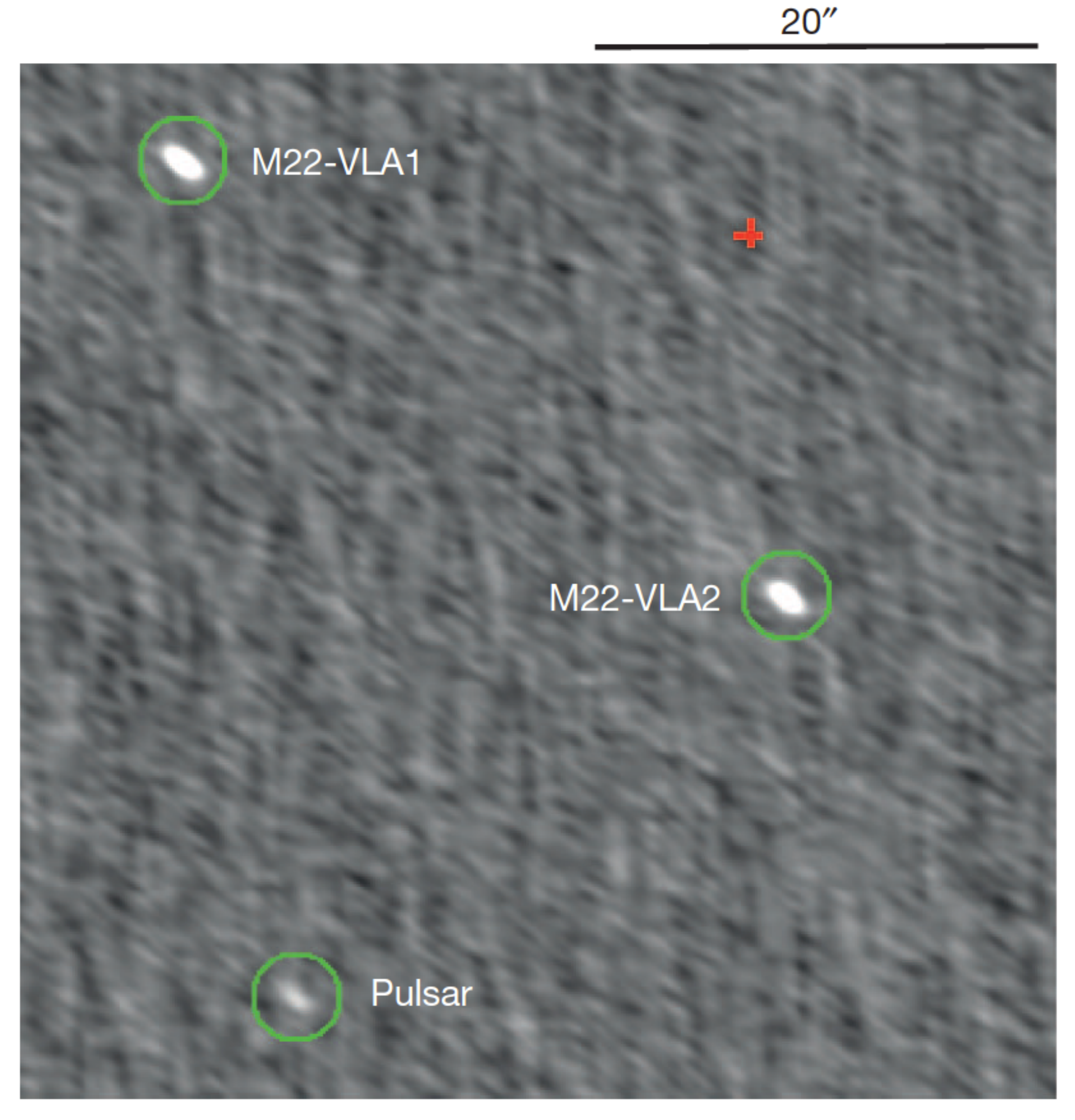}
\caption{VLA radio continuum image of the core of the globular cluster M22. The two bright circled objects are the sources identified as stellar-mass black holes, M22-VLA1 and M22-VLA2. These sources have flux densities of 55--58 $\mu$Jy at 5.9 GHz. We obtained the data in two separate 1 GHz basebands centered at 5 and 6.75 GHz, allowing a measurement of the spectral index of the radio emission between these frequencies. Both sources have flat radio spectra, with $\alpha = 0-0.2$, assuming $S_{\nu} = \nu^{\alpha}$. The faint circled object is a known millisecond pulsar\cite{Lyn11}. A red cross marks the photometric cluster center. $20''$ corresponds to approximately 0.3 pc at the distance of M22. The apparent elongation of the sources is due entirely to the elongated synthesized beam; all three circled sources are unresolved. North is up and east is to the left in this image.}
\end{figure}

\begin{figure}
\centering
\includegraphics[width=6.5in]{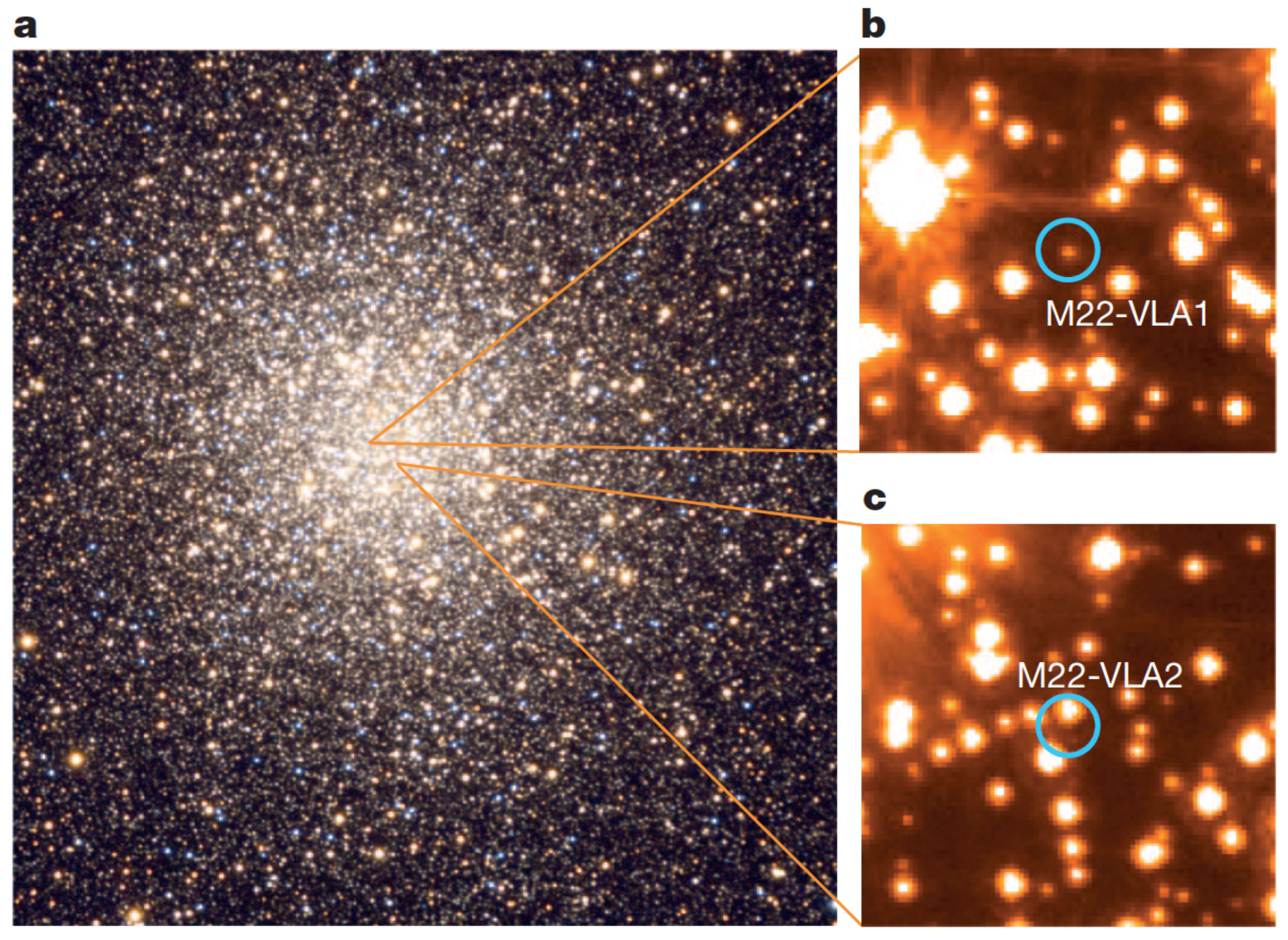}
\caption{Optical images of M22 and the candidate companion stars to the radio sources. (a) Ground-based image that shows the approximate location of the sources in the context of the star cluster. (b) and (c) show the zoomed-in location of the radio sources on an archival high-resolution \emph{Hubble Space Telescope}/Advanced Camera for Surveys $F814W$ image. These circles have radii of $0.3''$ for clarity; the uncertainty in the astrometric matching of the optical and radio data is $< 0.1''$. The image orientation is as in Fig. 1. (Image credit for (a): Doug Matthews/Adam Block/NOAO/AURA/NSF.)}
\end{figure}

\begin{figure}
\centering
\includegraphics[width=6in,angle=0]{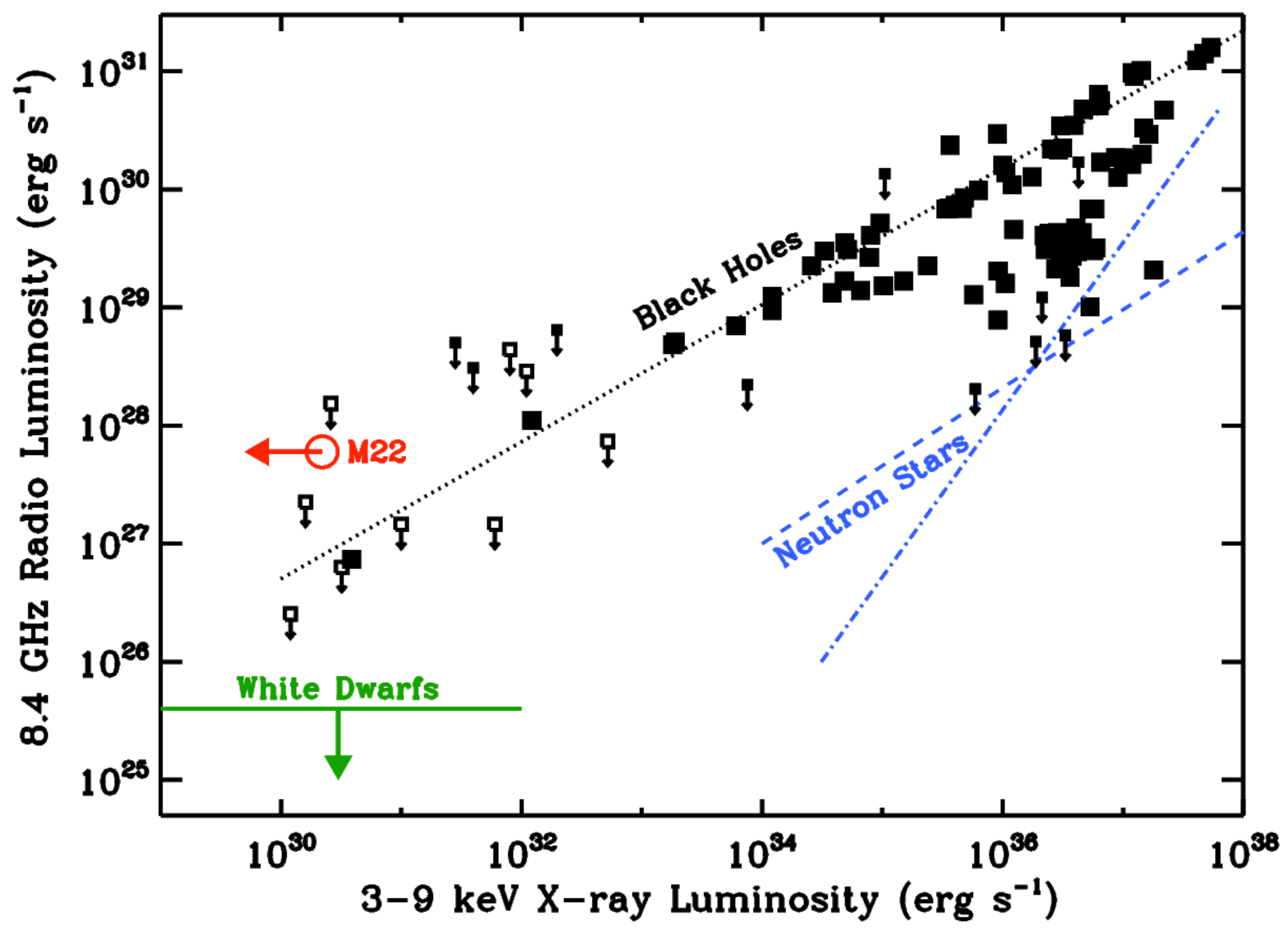}
\caption{Radio--X-ray correlation for stellar-mass black holes. The M22 sources have properties more consistent with black holes than neutron stars
or white dwarfs. Filled squares represent simultaneous radio and X-ray data; unfilled squares are non-simultaneous measurements for which variability may affect their positions. Upper limits are also shown. Some objects have multiple measurements plotted that represent different phases of accretion. The open red circle represents both M22-VLA1 and M22-VLA2, which have very similar luminosities. The dotted black line represents the published correlation\cite{Gal06} $L_R \propto L_X^{0.58}$, normalized by a least-squares fit to the simultaneous detections with $L_X < 2\times10^{34}$ erg s$^{-1}$. The dashed and dot-dashed blue lines show two possible radio--X-ray correlations for accreting neutron stars; this relation is poorly constrained by observations\cite{Mil06}. The solid green line shows the maximum radio continuum luminosity observed for accreting white dwarfs\cite{Kor11}. Neither neutron stars nor white dwarfs have properties consistent with the M22 radio sources. More information can be found in the Supplementary Information.}
\end{figure}

\clearpage
\pagestyle{empty}

{\bf {\Large Supplementary Information}}\\

\setlength{\parindent}{24pt} 

\section{Data Acquisition and Reduction}

We observed M22 with the VLA in May 2011 as part of the program 10C-109 (P.I.~L.~Chomiuk). Ten hours were spent observing the cluster, split amongst four 2.5 hour blocks, for a total of 7.5 hours on each source. We observed in C band with 2 GHz total bandwidth and four polarization products. Of the two basebands of 1 GHz width, one was centered at 5.0 GHz and the other at 6.75 GHz. The array was in BnA configuration, which is the normal B configuration with an extended northern arm to permit improved observations of southerly sources. This configuration gave a resolution of $1.53^{\prime\prime} \times 0.81^{\prime\prime}$ at our mean frequency of 5.9 GHz. The field of view (full-width at half power) of the VLA at this frequency is $\sim  7.7^{\prime}$ in diameter, compared to the half-mass diameter of 6.7$^{\prime}$ for M22\cite{McL05}.

We observed J1820-2528 as the secondary phase calibrator and J1407+2827 as the polarization leakage calibrator. 3C286 was used as an absolute flux density, bandpass, and polarization angle calibrator. The data were reduced using standard routines in AIPS. Weights were applied using TYAPL, and each 2.5-hour block was edited for bad data and interference, followed by calibration. For each individual calibrated baseband, the data were concatenated in the $uv$ plane and then imaged. Figure 1 is a deep coadded image of both basebands, obtained by smoothing the 6.75 GHz baseband to the resolution of the 5.0 GHz basebands and averaging these together. The rms sensitivity of this coadded image is 1.5 $\mu$Jy beam$^{-1}$, one of the deepest radio continuum images ever published.

We measure flux densities on the individual image for each baseband. The measured flux densities for the two sources are: M22-VLA1: $56.3\pm3.9$ $\mu$Jy (5.0 GHz), $60.1\pm3.6$ $\mu$Jy (6.75 GHz); M22-VLA2: $54.6\pm4.3$ $\mu$Jy (5.0 GHz), $54.5\pm3.5$ $\mu$Jy (6.75 GHz). If we assume the flux densities follow a power law of the form $S_{\nu} = \nu^{\alpha}$,  $\alpha=0.2\pm0.3$ and $0.0\pm0.3$ for sources M22-VLA1 and M22-VLA2 respectively. We set upper limits of $\sim 10$\% on both the linear and circular polarization of the sources at C band.

Matching of the VLA and $HST$ astrometric systems was required to search for optical counterparts to the radio sources. We fixed the astrometric zeropoint of the F814W ($I$-equivalent) $HST$ image to the VLA data through a match between an obvious background galaxy in the $HST$ image and a resolved continuum source in the VLA images. This mapping was confirmed to a precision $< 0.1''$ with the X-ray image, which also has sources in common with each of the $HST$ and VLA data. The J2000 positions of the sources are, in the reference frame of the VLA images: M22-VLA1 (18:36:25.825 --23:54:13.75), M22-VLA2 (18:36:23.824 --23:54:33.50). The formal uncertainties in these positions are $\sim 0.02-0.03''$.

\section{Calculation of X-ray Upper Limits}

The archival Chandra ACIS-S imaging has a total integration time of 16 kiloseconds. If we assume a completeness limit of 8 counts, then at the distance of M22 ($3.2\pm0.3$ kpc\cite{Mon04}) these data are sensitive to a luminosity of $L_X \sim 2.2 \times 10^{30}$ erg s$^{-1}$ over the range 3--9 keV assuming a foreground $N_H = 1.7\times10^{21}$ cm$^{-2}$ and a photon index of 2. If instead taken over the Chandra energy band of 0.5--8 keV, the limit is $\sim 5.6 \times 10^{30}$ erg s$^{-1}$.

\section{Photometry of Candidate Optical Counterparts}

The optical photometry of the candidate counterparts\cite{And08}, corrected for differential reddening\cite{Alo12}, is: $F606W_0 = 21.77\pm0.11$; $(F606W-F814W)_0 = 1.04\pm0.11$ (M22-VLA1) and $F606W_0 = 18.62\pm0.01$; $(F606W-F814W)_0 = 0.59\pm0.02$ (M22-VLA2). The listed photometric uncertainties do not include the uncertainty in the extinction. For a distance of 3.2 kpc, the inferred absolute magnitudes are $M_{F814W} = 8.21$ and 5.51 respectively. These values were converted into main sequence masses using a standard stellar isochrone for these parameters\cite{Dot07}: 13 Gyr, [Fe/H] = $-1.7$, [$\alpha$/Fe] = +0.2.

Compared to the ridge line of main sequence stars in M22, the candidate counterpart for M22-VLA1 lies $0.08\pm0.11$ mag blueward in $(F606W-F814W)$. A blue excess could be taken as evidence for hot emission from a putative accretion disk, but given the large error in the color the difference is not statistically significant.

\section{Other Candidates for the Sources}

We have discussed the classification of the radio continuum sources as stellar-mass black holes in the main text. We believe an unlikely (but possible) alternative classification is as background sources, which we consider first. Then we discuss other possible classifications that we believe are very unlikely or can be ruled out. Future very long baseline interferometry can be used to measure the proper motions of M22 VLA-1 and VLA-2 and definitively confirm or refute their association with M22.

\noindent
{\bf Background sources:} Very little information exists about the radio properties of background objects below 100 $\mu$Jy in flux density, especially when spectral index information is included. Using the limited deep background source counts from the literature\cite{Kel08}, we find that $\sim 0.2$--0.4 background objects with flat spectra and flux densities of 30--100 $\mu$Jy would be expected in the $1.3'$ radius core of M22. In the central $30''$, $< 0.1$ source would be expected. These calculations do not account for the lack of optical counterpart, which reduces the expected counts by a factor $\gtrsim 2-3$ (see below). For these reasons, we believe that it is very unlikely, but not impossible, that M22-VLA1 and M22-VLA2 are both background sources.

Background galaxies can have flat radio spectra through two scenarios: active galactic nuclei with partially self-absorbed jets, or star-forming galaxies in which the synchrotron emission is ``diluted" with thermal emission to produce a flat composite spectrum. Active galactic nuclei are expected to dominate at the flux densities of our sources. Most background radio continuum sources in the observed range of flux densities are found to have optical galaxy counterparts brighter than the $HST$ completeness limit of $I\sim24$\cite{Cil05,Pad09}. This is especially true of sources with flat radio spectra, many of which are associated with relatively bright early-type galaxies at $z < 1$\cite{Mai08}. There are no candidate optical background galaxies that are plausibly associated with either source. By contrast, in the outer regions of our M22 VLA image there is a canonical background galaxy: it has a flux density of $\sim 100 \mu$Jy at 5.9 GHz and is resolved in the VLA data, it has a spectral slope $\alpha = -0.8$, and has an obvious distant early-type galaxy as an optical counterpart in the $HST$ images.

Two published studies with multi-band radio continuum imaging can be used to assess the occurrence of background dopplegangers to the M22 sources. The first uses deep VLA 1.4 and 8.4 GHz data and $HST$ imaging\cite{Fom02}. Of a catalog of 65 sources with optical data, only two have inferred 5.9 GHz flux densities in the range 30 to 100 $\mu$Jy and $\alpha = -0.2$ to +0.2, and both have bright ($I < 20$) early-type galaxy counterparts. Another study used VLA 1.4 and 4.8 GHz data in the Chandra Deep Field South\cite{Kel08}. Here there are 12 sources that satisfy these flux density and slope ranges, though some of the spectral slopes have large uncertainties. Eight of 12 are bright enough that they would have been detectable in the $HST$ data; two others are borderline. From these (limited) comparisons we conclude that the lack of background optical counterparts reduces the expected rate of radio sources with the observed properties by at least a factor of 2--3.

\noindent
{\bf Pulsars/Supernova remnants:}
Millisecond pulsars are commonly found in globular clusters. However, these have universally steep radio spectra\cite{Kra99} ($\alpha \ltrsim -1$), inconsistent with the flat measured spectra. In rare cases pulsars drive a wind that interacts with ambient material, termed a pulsar wind nebula. These objects have flat measured spectra and $L_R/L_X$ consistent with the constraints for our sources. However, they also generally have $L_X > 10^{34}$ erg/s, a high degree of radio polarization, large sizes, are short lived, and are found in regions with dense gas, all inconsistent with observations\cite{Gae06}. Supernova remnants have similar properties and are also excluded.

\noindent
{\bf Accreting compact objects:}
A number of candidates involve accreting neutron stars or white dwarfs. For most companions these systems will have much lower $L_R/L_X$ or $L_R$ than accreting black holes, so are not consistent with the observed constraints on our sources\cite{Mig06,Kor11,Byc10,Fue86}. Figure 3 shows this graphically. The black hole data in this figure are compiled from recent papers on stellar-mass black holes\cite{Gal12, Mil11,Rat12}.

A contrived exception to this argument is if the sources are very strongly variable; the radio luminosities could be consistent with flares from accreting neutron stars, which then might have $L_X$ just below that which would have set off all-sky X-ray monitoring satellites. However, this scenario would require the coincidence of both neutron stars flaring simultaneously. In addition, they would need to be abnormally faint in quiescence\cite{Hei03}, since the typical $L_X $ for quiescent accreting neutron stars is $> 10^{32}$ erg s$^{-1}$. 

Symbiotic stars, which are white dwarfs accreting from luminous red giants, are ruled out by the optical source matching.

\noindent
{\bf Planetary Nebulae:}
Planetary nebulae can emit optically-thin thermal emission at radio wavelengths, and would therefore show a flat spectral index consistent with our sources. However, planetary nebulae are also bright [O III] $5007$\AA\ emission line sources. No [O III] nebulosity is observed around either source in archival $HST$/WFPC2 images obtained in the narrow F502N filter. By contrast, the known M22 planetary nebula IRAS 18333-2357 is clearly detected in this $HST$ [O III] image, and is not detected in our VLA radio continuum image.

\noindent
{\bf Foreground ultracool dwarfs:}
The last remaining possibility is that the sources are nearby ($< 100$ pc), cool (late M to early L) active dwarf stars in the foreground, which satisfy the $L_R/L_X$ constraints and which can have flat radio spectra\cite{Ber10}. The stars cannot be too nearby ($\ltrsim 50$ pc) or else they would be detectable in the $HST$ data, so they would need to be in a relatively narrow range of $\sim 50-100$ pc, the larger end of which would require strong flares (and be inconsistent with the lack of strong radio variability of our sources). Some ultracool dwarfs also show strong circular polarization\cite{Hal07}, which we do not observe. Cool dwarfs will be more luminous in the infrared, so we have consulted archival $K$ data from the Vista VVV survey\cite{Sai12}. These data rule out a cool dwarf identification with source M22-VLA1. The constraint for source M22-VLA2 is less strong as it is located near the diffraction spike from a bright star. Thus we cannot fully discount this possibility for M22-VLA2, but consider it to be unlikely because of the rarity of these sources.

\section{Variability}

We searched for variability in the radio sources by imaging each 2.5-hr data block separately and measuring their flux densities at each epoch. Fig.~S1 shows the results. Over the week of observations, M22-VLA1 appears constant within the uncertainties, while the flux density of M22-VLA2 appears to decrease monotonically by about 20\%. This overall trend is significant at the $\sim2.6\sigma$ level. Thus, the evidence for variability is only suggestive with the present data, and additional observations are necessary to determine if significant variability is present.\\

\section{The Large Core Radius of M22}

The core radius of M22 is $1.24^{+0.02}_{-0.04}$ pc, which is fifth-largest among luminous ($\gtrsim 2\times10^{5} L_{\odot}$) Milky Way globular clusters (it is third-largest, next to M14 and M53, if the anomalous clusters $\omega$ Cen and NGC 2419 are excluded)\cite{McL05}. A number of processes can lead to an increase in the core radius for globular clusters that contain black holes, and could help explain the large core radius of M22. The process dominant in some theoretical simulations\cite{Mac08} is the formation of one or more black hole--black hole binaries in the cluster core. Three-body interactions between one of these binaries and another black hole can eject the black hole from the core (but not the cluster entirely), and as this black hole sinks back to the center through dynamical friction it heats stars, removing them from the core. It is not clear what the necessary minimum population of black holes is for this process to be effective\cite{Mer04}.

It is also possible that binaries containing a black hole and another object, such as a main sequence star or white dwarf, can also heat the core through three-body scattering interactions. Black hole binaries with Roche lobe-overflowing counterparts can have binding energies ($\sim 10^{48}$--$10^{50}$ erg s$^{-1}$) comparable to the total binding energy of all of the single stars in the core of a globular cluster. However, these interactions may instead lead to stellar mergers\cite{Fre04}.

A possibility frequently discussed in the literature is that stellar interactions with an intermediate-mass black hole ($> 100$ $M_{\odot}$) can also lead to heating and core expansion\cite{Tre07, Mio07}. Deep radio data for M22\cite{Str12} have ruled out central intermediate-mass black holes with masses $> 360$ $M_{\odot}$ ($3\sigma$) but less massive black holes could still be present. These would have a mass ratio $< 6 \times 10^{-4}$ compared to the cluster mass and so it is unclear whether they could be solely responsible for a significant increase in the core radius, but some contribution is possible.

Finally, it is possible that the large core radius of M22 is unconnected to the presence of black holes. It has been proposed that the dynamical evolution of most globular clusters is less advanced than generally supposed, and that many clusters are still in an initial, slow phase of core contraction\cite{Fre08,Cha12}. In this case, the unusually large core radius of M22 might just reflect the initial conditions of the formation of the cluster.

\section{The Origin of Multiple Black Holes in M22}

Here we mention a few possible explanations for the presence of multiple black holes in M22. The discussion in the main text emphasized theoretical work suggesting that only one black hole or black hole--black hole binary will generally remain through long-term dynamical evolution of a globular cluster. A contrasting view is drawn from simulations in which the presence of black holes causes significant core expansion\cite{Mac08}. This expansion leads to a large reduction in the stellar density in the core and slows the black hole interaction rate. Under these circumstances, a significant fraction of the initial complement of black holes remain bound over timescales of several Gyr. These simulations were designed to match Magellanic Cloud star clusters with lower masses and larger core radii than M22 and so have central densities much lower (by a factor $> 10^3$) than M22 at the present day. Comparable simulations tailored to the properties of M22 can help constrain the formation of M22-VLA1 and M22-VLA2.

A speculative hypothesis for the existence of multiple black holes arises from the observation that M22 has modest spread in [Fe/H]\cite{Mar11}. [Fe/H] dispersions are common among massive globular clusters and may be due to self-enrichment\cite{Wil12}. It has also been proposed (for M22 and other massive clusters) that an [Fe/H] spread is due to the merger of two or more globular clusters that were each homogeneous in [Fe/H], but with different mean levels of enrichment\cite{Mar12}. Such mergers could occur either in dwarf galaxies with low velocity dispersions or if the clusters were in binary systems. If these putative mergers happened after the formation and dynamical evolution of black hole subsystems in globular clusters, then, even if each cluster originally had only one remaining black hole, multiple black holes could be present after the merger.

\pagestyle{empty}

\setcounter{figure}{0}
\renewcommand{\figurename}{Figure S1}
 
\clearpage
\begin{figure}
%\centering
\includegraphics[width=6in,angle=0]{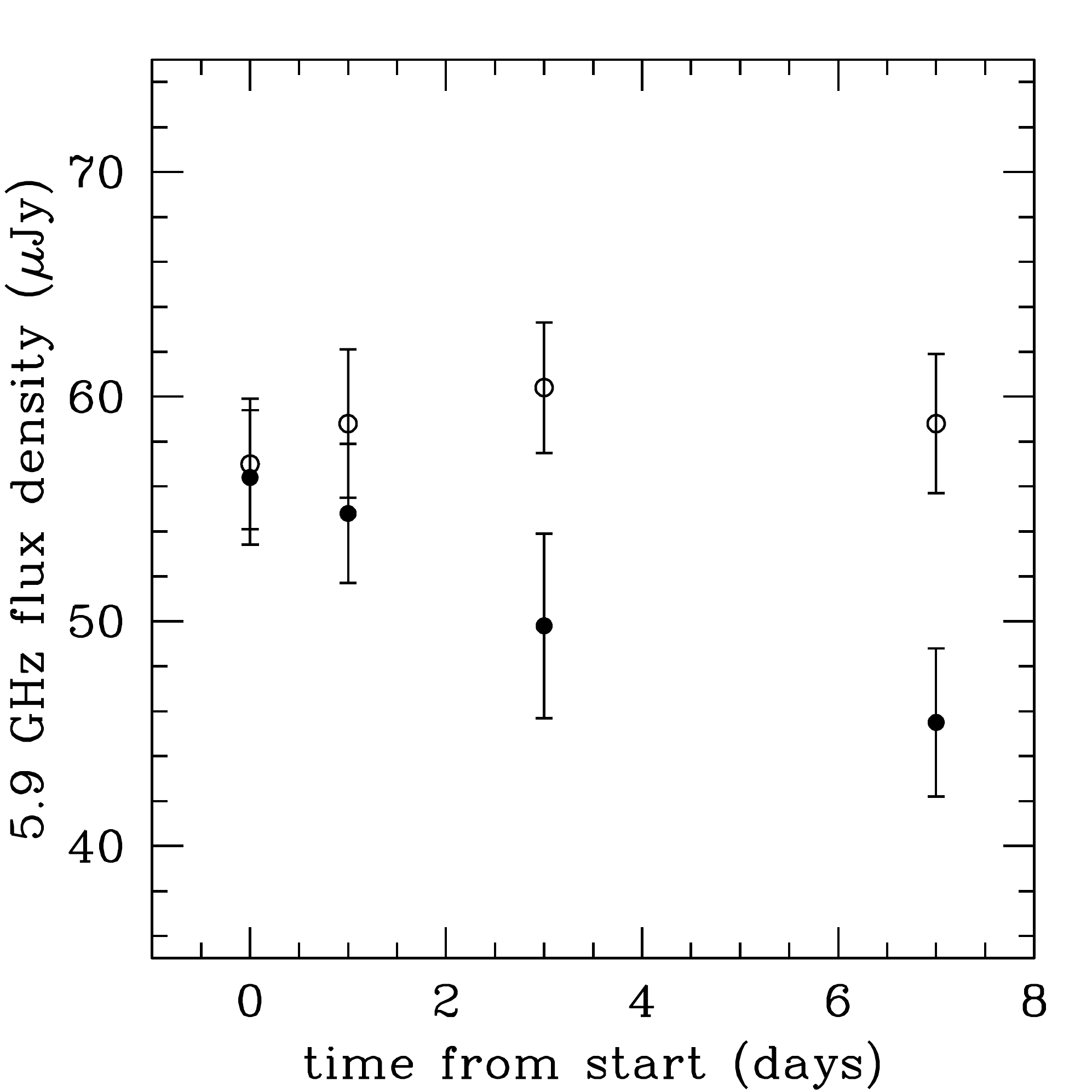}\\
{\bf Figure S1}: Time series of flux densities for M22 radio sources. M22-VLA1 (open circles) is constant over the week of observations,
while M22-VLA2 (filled circles) decreases over the same period at a significance of $2.6\sigma$. The error bars are standard deviations
of the measured flux densities.
\end{figure}


\section*{References}

\begin{thebibliography}{1}
\bibitem{Kul93} Kulkarni, S.~R., Hut, P. \& McMillan, S. Stellar black holes in globular clusters.\ {\it Nature}  {\bf 364}, 421-423 (1993)
\bibitem{Sig93} Sigurdsson, S. \& Hernquist, L. Primordial black holes in globular clusters.\ {\it Nature} {\bf 364}, 423-425 (1993)
\bibitem{Por00} Portegies Zwart, S.~F., \& McMillan, S.~L. W.  Black hole mergers in the universe.\ {\it Astrophys. J.} {\bf 528}, L17-L20 (2000) 
\bibitem{Kal04} Kalogera, V., King, A.~R., \& Rasio, F.~A. Could black hole X-Ray binaries be detected in globular clusters?  {\it Astrophys. J.} {\bf 601}, L171-L174 (2004)
\bibitem{Mac07a} Maccarone, T.~J., Kundu, A., Zepf, S.~E., \& Rhode, K.~L.  A black hole in a globular cluster.\ {\it Nature} {\bf 445}, 183-185 (2007)
\bibitem{Irw10} Irwin, J.~A., Brink, T.~G., Bregman, J.~N., \& Roberts, T.~P.  Evidence for a stellar disruption by an intermediate-mass black hole in an extragalactic globular cluster.\ {\it Astrophys. J.} {\bf 712}, L1-L4 (2010)
\bibitem{Por04} Portegies Zwart, S.~F., Baumgardt, H., Hut, P., Makino, J., \& McMillan, S.~L.~W.  Formation of massive black holes through runaway collisions in dense young star clusters.\ {\it Nature} {\bf 428}. 724-726 (2004)
\bibitem{Bel10} Belczynski, K. et al.\ On the maximum mass of stellar black holes.  {\it Astrophys. J.} {\bf 714}, 1217-1226 (2010)
\bibitem{Mac08} Mackey, A.~D., Wilkinson, M.~I., Davies, M.~B., \& Gilmore, G.~F.  Black holes and core expansion in massive star clusters.\ {\it Mon.~Not.~R.~Astron.~Soc}. {\bf 386}, 65-95 (2008)
\bibitem{Str12} Strader, J., et al.\ No evidence for intermediate-mass black holes in globular clusters: strong constraints from the VLA.
{\it Astrophys. J.} {\bf 750} L27 (2012)
\bibitem{McL05} McLaughlin, D. \& van der Marel, R. Resolved massive star clusters in the Milky Way and Its satellites: brightness profiles and a catalog of fundamental parameters. {\it Astrophys. J. Supp.} {\bf 161} 304-360 (2005)
\bibitem{Mon04} Monaco, L., Pancino, E., Ferraro, F.~R., \& Bellazzini, M. Wide-field photometry of the Galactic globular cluster M22.\ {\it Mon.~Not.~R.~Astron.~Soc}. {\bf 349} 1278-1290 (2004)
\bibitem{Mac07b} Maccarone, T., \& Knigge, C.\ Compact objects in globular clusters. {\it Astron. Geophys.} {\bf 48} 5.12-5.20 (2007)
\bibitem{Gal05} Gallo, E., Fender, R.~P., \& Hynes, R.~I.\ The radio spectrum of a quiescent stellar mass black hole. {\it Mon.~Not.~R.~Astron.~Soc}. {\bf 356}, 1017-1021 (2005)
\bibitem{And08} Anderson, J. et al.\ The ACS survey of Galactic globular clusters. V. Generating a comprehensive star catalog for each cluster.\ {\it Astron. J.} {\bf 135} 2055-2073 (2008)
\bibitem{Mac03} Maccarone, T.\ Do X-ray binary spectral state transition luminosities vary? {\it Astron. Astrophys.} {\bf 409} 697-706 (2003)
\bibitem{Gal03} Gallo, E., Fender, R.~P., \& Pooley, G.~G.\ A universal radio-X-ray correlation in low/hard state black hole binaries. {\it Mon.~Not.~R.~Astron.~Soc}. {\bf 344} 60-72 (2003)
\bibitem{Gal06} Gallo, E., et al.\ A radio-emitting outflow in the quiescent state of A0620-00: implications for modelling low-luminosity black hole binaries. {\it Mon.~Not.~R.~Astron.~Soc}. {\bf 370} 1351-1360 (2006)
\bibitem{Mil11} Miller-Jones, J.~C.~A., Jonker, P.~G., Maccarone, T.~J., Nelemans, G., \& Calvelo, D.~E.\ A deep radio survey of hard state and quiescent black hole X-Ray binaries. {\it Astrophys. J.} {\bf 739} L18 (2011)
\bibitem{Mil08} Miller-Jones, J.~C.~A, et al.\ Zooming in on a sleeping giant: milliarcsecond High Sensitivity Array imaging of the black hole binary V404 Cyg in quiescence. {\it Mon.~Not.~R.~Astron.~Soc}. {\bf 388} 1751-1758 (2008)
\bibitem{Cor06} Corbel, S., Tomsick, J.~A., \& Kaaret, P.\ On the origin of black hole X-Ray emission in quiescence: Chandra observations of XTE J1550-564 and H1743-322. {\it Astrophys. J.} {\bf 636} 971-978 (2006)
\bibitem{Mer03}  Merloni, A., Heinz, S., \& di Matteo, T.\ A fundamental plane of black hole activity. {\it Mon.~Not.~R.~Astron.~Soc}. {\bf 345} 1057-1076 (2003)
\bibitem{Rem06} Remillard, R.~A., \& McClintock, J.~E.\ X-Ray properties of black-hole binaries. {\it Ann. Rev. Astron. Astrophys.} {\bf 44} 49-92 (2006)
\bibitem{Tau06} Tauris, T.~M. \& van den Heuvel, E.~P.~J. in {\it Compact Stellar X-ray Sources} (eds. W.~H.~G.~Lewin \& M. van der Klis) 623-665 (Cambridge University Press, Cambridge, 2006)
\bibitem{Iva10} Ivanova, N., et al.\ Formation of black hole X-ray binaries in globular clusters. {\it  Astrophys. J.} {\bf 717} 948-957 (2010)
\bibitem{Rep12} Repetto, S., Davies, M.~B., Sigurdsson, S. Investigating stellar-mass black hole kicks. {\it Mon.~Not.~R.~Astron.~Soc}, in press (2012) {\tt arXiv:1203.3077}
\bibitem{Lyn11} Lynch, R.~S., Ransom, S.~M., Freire, P.~C.~C., \& Stairs, I.~H.\ Six new recycled globular cluster pulsars discovered with the Green Bank Telescope. {\it Astrophys. J.} {\bf 734} 89 (2011)
\bibitem{Mil06} Migliari, S., \& Fender, R.~P.\ Jets in neutron star X-ray binaries: a comparison with black holes. {\it  Mon.~Not.~R.~Astron.~Soc}. {\bf 366} 79-91 (2006)
\bibitem{Kor11}  K{\"o}rding, E.~G., Knigge, C., Tzioumis, T., \& Fender, R.\ Detection of radio emission from a nova-like cataclysmic variable: evidence of jets? {\it Mon.~Not.~R.~Astron.~Soc}. {\bf 418} L129-L132 (2011)
%\bibitem{Osh10} O'Shaughnessy, R., Kalogera, V., \& Belczynski, K.\ Binary compact object coalescence rates: The role of elliptical galaxies. {\it Astrophys. J.} {\bf 715}, 615-633 (2010)
%\bibitem{}\\

\pagestyle{empty}

\noindent
{\bf References}
\bibitem{Alo12} Alonso-Garc{\'{\i}}a, J. et al.\ Uncloaking globular clusters in the inner Galaxy. {\it Astron. J.} {\bf 143} 70 (2012)
\bibitem{Dot07} Dotter, A. et al.\ The ACS survey of Galactic globular clusters. II. stellar evolution tracks, isochrones, luminosity functions, and synthetic horizontal-branch models. {\it Astron. J.} {\bf 134} 376-390 (2007)
\bibitem{Kel08} Kellermann, K.~I. et al.\ The VLA survey of the Chandra Deep Field-South. I. Overview and the radio data. {\it Astrophys. J. Supp.} {\bf 179} 71-94 (2008)
\bibitem{Cil05} Ciliegi, P. et al.\ The VVDS-VLA deep field. II. Optical and near infrared identifications of VLA S$_{\rm 1.4 GHz}$ $> 80 \mu$Jy sources in the VIMOS VLT deep survey VVDS-02h field. {\it Astron. Astrophys.} {\bf 441} 879-891 (2005)
\bibitem{Pad09} Padovani, P. et al.\ The Very Large Array survey of the Chandra Deep Field South. IV. Source population. {\it Astrophys. J.} {\bf 694} 235-246 (2009)
\bibitem{Mai08} Mainieri, V., et al.\ The VLA survey of the Chandra deep field-south. II. Identification and host galaxy properties of submillijansky sources. {\it Astrophys. J. Supp.} {\bf 179} 95-113 (2008)
\bibitem{Fom02} Fomalont, E.~B., Kellermann, K.~I., Partridge, R.~B., Windhorst, R.~A., \& Richards, E.~A.\ The microjansky sky at 8.4 GHz. {\it Astron. J.} {\bf 123} 2402-2416 (2002)
\bibitem{Kra99} Kramer, M., et al.\ The characteristics of millisecond pulsar emission. III. From low to high frequencies. {\it Astrophys. J.} {\bf 526} 957-975 (1999)
\bibitem{Gae06} Gaensler, B.~M., \& Slane, P.~O.\ The evolution and structure of pulsar wind nebulae. {\it Ann. Rev. Astron. Astrophys.} {\bf 44} 17-47 (2006)
\bibitem{Mig06} Migliari, S., \& Fender, R.~P.\ Jets in neutron star X-ray binaries: a comparison with black holes. {\it Mon.~Not.~R.~Astron.~Soc}. {\bf 366} 79-91 (2006)
\bibitem{Byc10}  Byckling, K., Mukai, K., Thorstensen, J.~R., \& Osborne, J.~P.\ Deriving an X-ray luminosity function of dwarf novae based on parallax measurements.  {\it Mon.~Not.~R.~Astron.~Soc}. {\bf 408} 2298-2311 (2010)
\bibitem{Fue86} Fuerst, E., Benz, A., Hirth, W., Geffert, M., \& Kiplinger, A.\ Radio emission of cataclysmic variable stars. {\it Astron. Astrophys.} {\bf 154} 377-378 (1986)
\bibitem{Gal12} Gallo, E., Miller, B., Fender, R.~P. Assessing luminosity correlations via cluster analysis: evidence for dual tracks in the radio/X-ray domain of black hole X-ray binaries. {\it Mon.~Not.~R.~Astron.~Soc}. {\bf 423} 590-599 (2012)
\bibitem{Rat12} Ratti, E.~M., et al.\ The black hole candidate XTE J1752-223 towards and in quiescence: optical and simultaneous X-ray--radio observations. {\it Mon.~Not.~R.~Astron.~Soc}. {\bf 423} 2656-2667 (2012)
\bibitem{Hei03} Heinke, C.~O., et al.\ Analysis of the quiescent low-mass X-Ray binary population in Galactic globular clusters. {\it Astrophys. J.} {\bf 598} 501-515 (2003)
\bibitem{Ber10} Berger, E., et al.\ Simultaneous multi-wavelength observations of magnetic activity in ultracool dwarfs. III. X-ray, radio, and H$\alpha$ activity trends in M and L dwarfs. {\it Astrophys. J.} {\bf 709} 332-341 (2010)
\bibitem{Hal07} Hallinan, G., et al.\ Periodic bursts of coherent radio emission from an ultracool dwarf. {\it Astrophys. J.} {\bf 663} L25-L28 (2007)
\bibitem{Sai12} Saito, R.~K., et al.\ VVV DR1: The first data release of the Milky Way bulge and southern plane from the near-infrared ESO public survey VISTA variables in the Via Lactea. {\it Astron. Astrophys.} {\bf 537} A107 (2012)
\bibitem{Mer04} Merritt, D., Piatek, S., Portegies Zwart, S., \& Hemsendorf, M.\ Core formation by a population of massive remnants. {\it Astrophys. J.} {\bf 608} L25-L28 (2004)
\bibitem{Fre04} Fregeau, J., Cheung, P., Portegies Zwart, S.~F., \& Rasio F.~A.\ Stellar collisions during binary-binary and binary-single star interactions. {\it Mon.~Not.~R.~Astron.~Soc}. {\bf 352} 1-19 (2004)
\bibitem{Tre07} Trenti, M., Ardi, E., Mineshige, S., \& Hut, P.\ Star clusters with primordial binaries - III. Dynamical interaction between binaries and an intermediate-mass black hole. {\it Mon.~Not.~R.~Astron.~Soc}. {\bf 374} 857-866 (2007)
\bibitem{Mio07} Miocchi, P.\ The presence of intermediate-mass black holes in globular clusters and their connection with extreme horizontal branch stars. {\it Mon.~Not.~R.~Astron.~Soc}. {\bf 381} 103-116 (2007)
\bibitem{Fre08} Fregeau, J.\ X-Ray binaries and the current dynamical states of Galactic globular clusters. {\it Astrophys. J.} {\bf 673} L25-L28 (2008)
\bibitem{Cha12} Chatterjee, S., Umbreit, S., Fregeau, J., \& Rasio F.~A.\ Understanding the dynamical state of globular clusters: Core-collapsed vs.~non core-collapsed. {\it Mon.~Not.~R.~Astron.~Soc}, submitted (2012) {\tt arXiv:1207.3063}
\bibitem{Mar11} Marino, A., et al.\ The two metallicity groups of the globular cluster M 22: A chemical perspective. {\it Astron. Astrophys.} {\bf 532} A8 (2012)
\bibitem{Wil12} Willman, B., \& Strader, J.\ ``Galaxy," defined. {\it Astron. J.} {\bf 144} 76 (2012)
\bibitem{Mar12} Marino, A., et al.\ The double sub-giant branch of NGC 6656 (M 22): A chemical characterization. {\it Astron. Astrophys.} {\bf 541} A15 (2012)

\end{thebibliography}
\end{document}